\documentclass[lettersize,journal]{IEEEtran}
\usepackage{amsmath,amsfonts}
\usepackage{algorithmic}
\usepackage{algorithm}
\usepackage{array}
\usepackage[caption=false,font=normalsize,labelfont=sf,textfont=sf]{subfig}
\usepackage{textcomp}
\usepackage{stfloats}
\usepackage{url}
\usepackage{verbatim}
\usepackage{graphicx}
\usepackage{cite}
\usepackage{color}
\usepackage{upgreek}
\usepackage{amssymb}

\makeatletter

\newcommand{\Rmnum}[1]{\expandafter\@slowromancap\romannumeral #1@}
\makeatother

\begin{document}

\title{AMDET: Attention based Multiple Dimensions EEG Transformer for Emotion Recognition}

\author{Yongling Xu,
        ~Yang Du,
        ~Jing Zou,
        ~Tianying Zhou,
        ~Lushan Xiao,
        ~Li Liu,
        ~and Pengcheng Ma
\thanks{Yongling Xu and Yang Du contributed equally to this work. Corresponding authors: Lushan Xiao; Li Liu; Pengcheng Ma.}
\thanks{Yongling Xu and Pengcheng Ma are with the Big Data Center, Nanfang Hospital, Southern Medical University, Guangzhou, 510515, China, and also with the Brainup Research Lab, Naolu Technology Co., Ltd., Beijing, 100124, China (e-mail: pc.ma@foxmail.com).}
\thanks{Yang Du and Li Liu are with the Big Data Center, Nanfang Hospital, Southern Medical University, Guangzhou, 510515, China (e-mail: liuli@i.smu.edu.cn).}
\thanks{Jing Zou is with the Department of Psychology and Economics, University of California, Los Angeles, Los Angeles, CA 90024 USA.}
\thanks{Tianying Zhou is with the Department of Electrical Engineering and Computer Sciences, UC Berkeley, Berkeley, CA 94720 USA.}
\thanks{Lushan Xiao is with the Department of Infectious Diseases, Nanfang Hospital, Southern Medical University, Guangzhou, 510515, China (e-mail: 15622178423@163.com).}}



\maketitle

\begin{abstract}
Affective computing is an important subfield of artificial intelligence, and with the rapid development of brain-computer interface technology, emotion recognition based on EEG signals has received broad attention. It is still a great challenge to effectively explore the multi-dimensional information in the EEG data in spite of a large number of deep learning methods . In this paper, we propose a deep model called Attention-based Multiple Dimensions EEG Transformer (AMDET), which can leverage the complementarity among the spectral-spatial-temporal features of EEG data by employing the multi-dimensional global attention mechanism. We transformed the original EEG data into 3D temporal-spectral-spatial representations and then the AMDET would use spectral-spatial transformer encoder layer to extract effective features in the EEG signal and concentrate on the critical time frame with a temporal attention layer. We conduct extensive experiments on the DEAP, SEED, and SEED-IV datasets to evaluate the performance of AMDET and the results outperform the state-of-the-art baseline on three datasets. Accuracy rates of 97.48\%, 96.85\%, 97.17\%, 87.32\% were achieved in the DEAP-Arousal, DEAP-Valence, SEED, and SEED-IV datasets, respectively. We also conduct extensive experiments to explore the possible brain regions that influence emotions and the coupling of EEG signals. Remarkably, AMDET can perform as well even with few channels which are identified by visualizing what the trained model learned. The accuracy could achieve over 90\% even with only eight channels and it is of great use and benefit for practical applications. 
\end{abstract}


\section{Introduction}
Emotion is a comprehensive psychological and physiological response of human beings to an external event or stimulus. It can greatly impact a person's behavior and thoughts, and in some cases even affect health and lead to illness. There is no doubt that emotions play an important role in life. Thus, on a matter of such significance, emotion recognition technology has been widely introduced and used in life, such as mental illness detection, fatigue driving detection, and human-computer interaction. Therefore, more and more researchers have devoted to this research. 

Emotion recognition can be broadly classified into two categories, one of which is based on human external responses, such as facial expressions\cite{tarnowski2017emotion}, gestures\cite{noroozi2018survey} and voice intonation\cite{alu2017voice}, etc. The other is based on human physiological signals, such as breathing, heart rate, body temperature, electroencephalogram (EEG) and so on\cite{dzedzickis2020human}. The former is easier to collect but more subjective. To elaborate, people may fake their facial expressions and behavior, or deliberately speak loudly to pretend they are angry. Even in the same mood, different people behave differently. In contrast, EEG detection is more objective because humans struggle to control their physiological signals to fake emotions.

EEG is the bioelectric activity detected on the surface of the human scalp. It can be collected by portable and relatively inexpensive devices. The amplitude of EEG signal in normal humans ranges from 10$\upmu$V-200$\upmu$V\cite{he2020electrophysiological}, the frequency is 0.2Hz-90Hz\cite{biasiucci2019electroencephalography}. EEG has a high temporal resolution, which allows for the recording of brain activity with a resolution of milliseconds. However, it is limited in its spatial resolution, which refers to the ability to accurately locate brain activity within specific regions of the brain. This limitation is caused by the physical constraints of the EEG collection device, as well as the interference of the electric field between different areas of the brain. Despite these limitations, EEG remains a valuable tool for studying brain activity and has contributed significantly to our understanding of the brain.

Numerous studies have shown that EEG has the ability to accurately reflect an individual's emotional state to a certain degree. The characteristics of EEG in the time domain, space domain, and frequency domain are all highly correlated with human emotional states. For example, in the frequency domain, alpha waves are enhanced when people are in a calm state, beta waves are intensified when the brain is active and highly focused, and gamma waves are associated with hyperactivity in the brain\cite{wang2021review}. Researchers often use the ratio of beta waves to alpha waves as an indication of brain activeness\cite{murata1994quantitative,ray1985eeg}, and assess whether a person is in a happy state according to the power of theta waves\cite{sammler2007music}. Therefore, the power characteristics such as power spectral density (PSD)\cite{yin2020locally} and differential entropy (DE)\cite{duan2013differential,song2018eeg}, are often used as a feature of the EEG signal in many studies of emotion recognition\cite{cao2019research}.
\IEEEpubidadjcol

The spatial properties of EEG are reflected in the close correlation between each emotional state and some specific areas of the brain. The brain can be divided into four areas, which are the frontal lobe, which are the frontal lobe, parietal lobe, temporal lobe, and occipital lobe. The main functions of the frontal lobe are cognitive thinking and emotional demands. The parietal lobe responds to the tactile sense. It is also related to the body’s balance and coordination. The temporal lobe is mainly responsible for auditory and olfactory sensations as well as associated with emotional and mental activity. Finally, the occipital lobe is in charge of processing visual information\cite{wang2021review}. People's emotions would trigger activities in specific brain areas. For example, the activity of the left frontal lobe of the brain is activated when people feel happy\cite{ekman1993voluntary} and suppressed when people feel fearful\cite{bhatti2016human}. Li et al. combined functional connectivity networks with local activation to validate the activities of local brain regions responsive to emotions and the interaction between the brain regions involved in the activities\cite{li2019eeg}. It is evident that the EEG signal holds promising characteristics that can be examined in the spatial domain.

EEG has a high temporal resolution and therefore contains a lot of information in the time domain which should not be neglected. Some studies use statistic features to analyze the EEG signal, such as calculating the mean, standard deviation and difference etc. over a time window. These characteristics can indicate whether the EEG has oscillated smoothly or changed drastically during the time window period. In the field of emotion recognition, EEG's the first difference (1ST) is commonly used as a feature, which is defined as the mean of absolute values of the first second of the raw signal\cite{zheng2021portable}. Stationarity is also worth being considered when analyzing sequential signals. One of the measurements is the Lyapunov exponents, which are used to determine the stability of any steady-state behavior, including chaotic solutions\cite{guler2005recurrent}. Fourier transform is a common way to analyze the frequency domain. However, it does not exhibit any time-domain characteristic, so researchers have proposed the Short Time Fourier Transform to compensate for this defect. Wavelet transform is another solution to the lack of time domain information, which is also commonly applied to analyze EEG. It employs wavelets of different scales to model the signal that maximizes the preservation of time-domain information. These extended transform methods precisely illustrate the necessity of the EEG temporal features.

As mentioned above, EEG contains a potential abundance of information in the frequency, space and time domains. Therefore, how to extract and make full use of this information becomes the greatest challenge.
The main contributions of this paper are described below.
\begin{enumerate}
    \item {We proposed a model named AMDET which is excellent at extracting features of EEG signals by employing the multi-dimensional global attention mechanism. AMDET outperforms other state-of-the-art methods on DEAP, SEED, and SEED-IV datasets. We also conducted an ablation experiment to demonstrate the necessity to use all the information in the three domains of time, space, and frequency.}
    \item {We conducted extensive visualization experiments using a Grad-CAM based algorithm to reveal the focus of the model on channels and figure out the brain region that contributes more to emotion recognition.}
    \item {We further investigated the redundancy of EEG signals by reducing the number of channels in the experiment. And we also validate the effectiveness of our model with only a few EEG channels. AMDET can achieve high performance even when using less than 20 percent of the EEG channels, which offers the possibility of practical applications.}
\end{enumerate}

The remainder of this article is organized as follows. Related works are described in Section \Rmnum{2}. Then, Section \Rmnum{3} introduces the details of the proposed AMDET, including EEG signal preprocessing, multidimensional features extraction and the classification algorithm. The experiments presented in Section \Rmnum{4} are designed to prove the effectiveness of the proposed AMDET. Section \Rmnum{5} shows the experiment results and discussion. Section \Rmnum{6} draws conclusions and future work.

\section{Related Work}
The current EEG-based emotion recognition will be divided into two main methods. One of them is to extract distinguishable features first, and then use the traditional machine learning method for classification. Another is to use the end-to-end deep learning method, which completes feature extraction and classification simultaneously. Deep learning has outperformed traditional machine learning methods in some areas, such as computer vision and natural language processing. 

Atkinson et al. combined statistical-based feature selection methods and support vector machine (SVM) emotion classifiers and achieved decent results\cite{atkinson2016improving}. Wavelet transform is a widely used method for time-frequency domain analysis as well as feature extraction of EEG\cite{subasi2007eeg}. Li et al. used Discrete Wavelet Transform to divide EEG signals into four frequency bands and calculated their entropy and energy as the features of the k-nearest neighbor classifier (KNN)\cite{li2018emotion}. Subasi et al. also employed Tunable Q Wavelet Transform (TQWT) as a feature extractor and then exercised rotation forest ensemble as a classifier, which utilized different classification algorithms such as KNN, SVM, artificial neural network, random forest, and other four different types of decision tree algorithms\cite{subasi2021eeg}.

It is difficult to find representative and valid features in complex cognitive processes due to the great differences among subjects. Compared to traditional machine learning algorithms, deep learning does not require prior knowledge and manual feature extraction allowing it to directly extract features from complex data. Within Deep Learning, Convolution Neural Networks (CNN) can extract local characteristics of the data, recurrent neural networks (RNN) excel at extracting information from time-series data, and Transformer focuses its attention on the more influential parts of the data. Designing the network structure for the model to fully extract the information from the EEG signal is a critical step. Du et al. employed self-attention mechanism in time and space domains to extract the critical EEG features\cite{du2022eeg}. Li et al. first extracted the DE features of each channel and then arranged these features into a two-dimensional signal according to their position on the brain surface, then utilized a hierarchical convolutional neural network (HCNN) to extract and classify the spatial representation\cite{li2018hierarchical}. In some research, Long-Short Term Memory (LSTM) was used to learn temporal features from EEG signals\cite{alhagry2017emotion}. Tao et al. introduced the self-attention mechanism into their network model to assign weights to each channel and used CNN and RNN to obtain the time-domain and space-domain features of EEG, respectively, and finally achieved good results\cite{tao2020eeg}. Jia et al. designed a 3D attention mechanism to realize the complementarity among the spatial-spectral-temporal features and discriminative local patterns in all features\cite{jia2020sst}. Xiao et al. proposed a four-dimensional attention-based neural network, which fuses information on different domains and captures discriminative patterns in EEG signals\cite{xiao20224d}.

\section{Method}
To fully capture the EEG signals’ abundant information in the frequency, space, and time domains, we introduce the global attention mechanism\cite{vaswani2017attention} into our model. Fig. 1 shows an overview of AMDET. It contains a spatial attention block, a spectral attention block, a temporal attention block, and a classification layer. The preprocessing of the EEG signal will be first introduced.

\begin{figure*}[!t]
\centering
\includegraphics[width=7in]{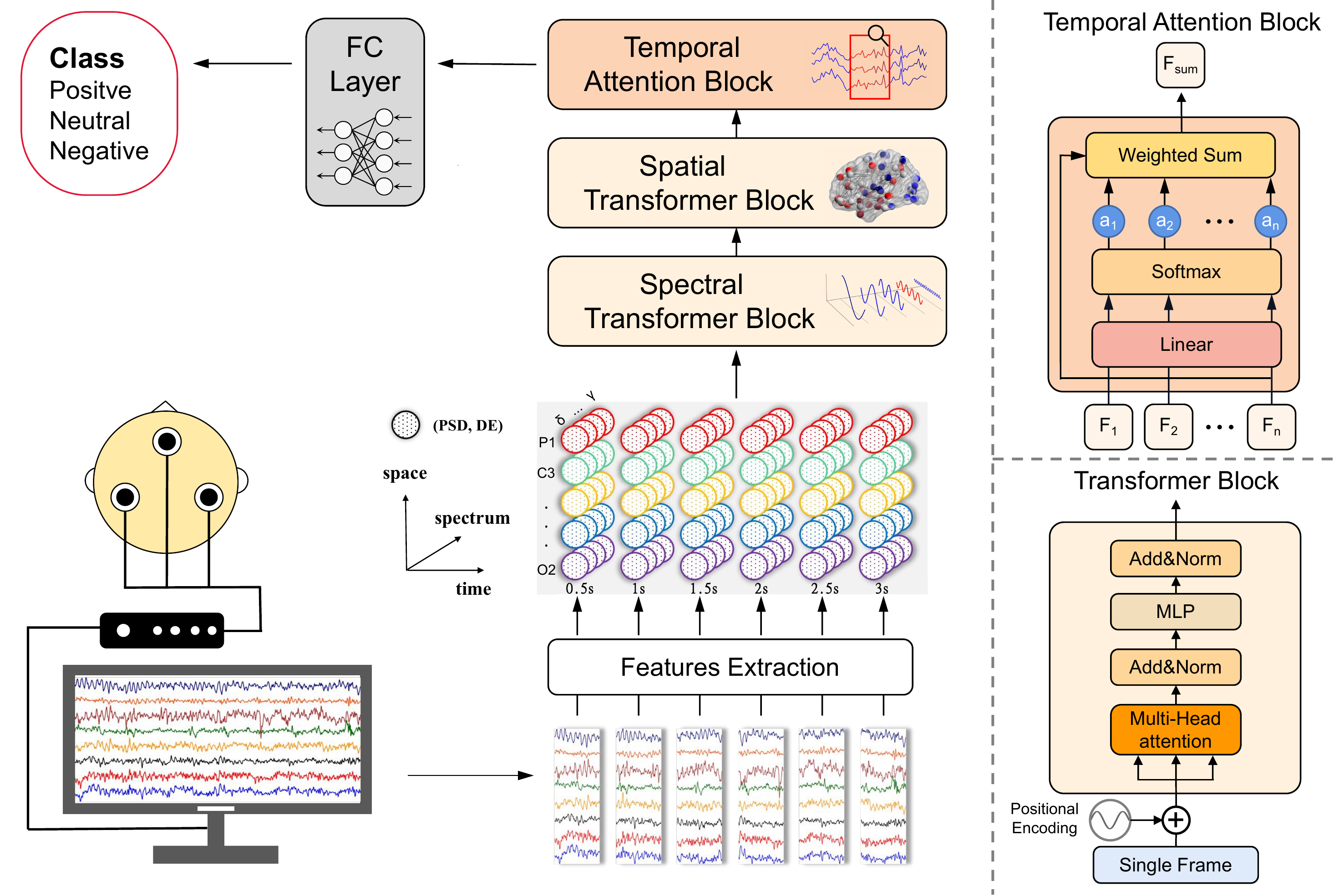}
\caption{The framework of our AMDET model for EEG emotion recognition, which consists of a spectral transformer block, a spatial transformer block, a temporal attention block, and a fully connected (FC) layer. The inputs of the model are 3D tensor containing differential entropy (DE) and power spectral density (PSD) extracted from each EEG channel and different time segments in multiple frequency bands including theta[4-8Hz], alpha[8-14Hz], beta[14-31Hz], gamma1[31-50Hz], and gamma2[50-75Hz]. The spectral transformer block and the spatial transformer block are used to discover and focus on the significant part of the input tensor in spectral and spatial domain separately. Similarly, the temporal attention block would aggregate all the frames and figure out the critical frame. Finally, a FC layer is used for classification. The figure description is made by the toolkit of BrainNet Viewer\cite{xia2013brainnet}}\label{fig:1}
\end{figure*}

\begin{figure}[ht]
\centering
\includegraphics[width=3.4in]{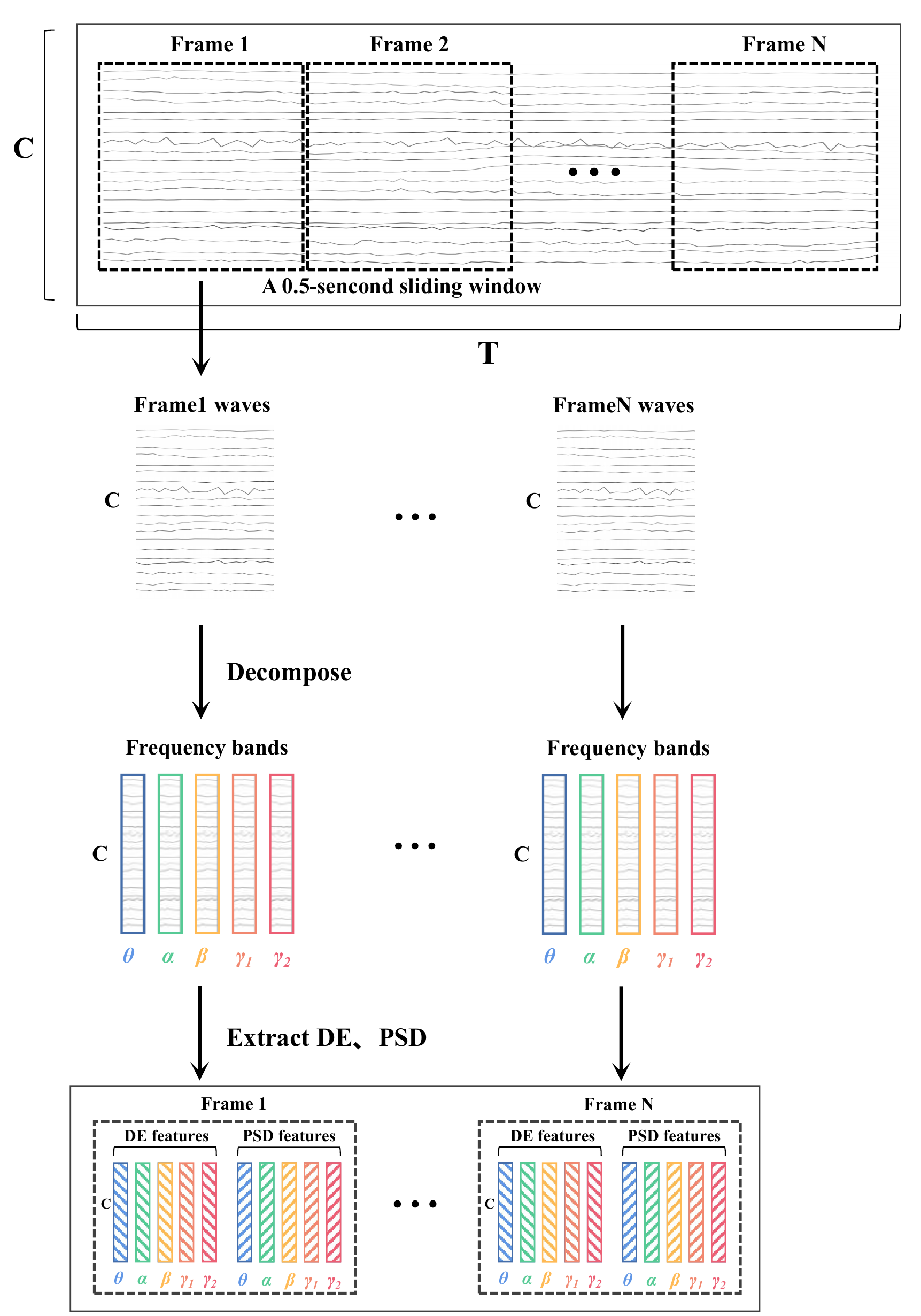}
\caption{The pre-processing of original EEG signals and the generation of the 3D tensor. We segments T seconds original EEG signals into N frames with a 0.5-second non-overlap sliding window. For each frame, we decompose it into five frequency bands according to fourier transform firstly, and then extract DE and PSD features from each frequency band and C channels.}\label{fig:2}
\end{figure}

\subsection{Pre-processing}
As shown in Fig. 2, the original EEG data need to be preprocessed to generate EEG samples with three dimensions. To exploit the time-domain information of the EEG data, a 3-second window is first used to perform a non-overlapping segmentation of the raw EEG data to generate samples. Then, each sample is divided without overlap using a 0.5-second window, and the EEG signal in each window is considered as a frame in the sample. A single EEG sample contains multiple consecutive frames to preserve the time-domain characteristics of the EEG signal. In addition, since EEG signals are collected from multiple channels, and different channels represent different brain regions, we reserve the EEG channels to retain spatial information. Moreover, it has been shown that the high-frequency part of the EEG signal has a greater effect on emotion recognition than the other frequency parts.\cite{li2009emotion}. Therefore, we divide the EEG signal into multiple frequency bands and extract features on each band. For each frame in a sample, it is filtered on theta[4-8Hz], alpha[8-14Hz], beta[14-31Hz], gamma1[31-50Hz], gamma2[50-75Hz], respectively. Since PSD and DE features have been shown to be effective in EEG emotion identification\cite{zheng2015investigating,duan2013differential}, we extract PSD and DE features for each frame on all five frequency bands for each channel separately. The PSD is defined as:
\begin{equation}
\label{equ_1}
PSD=E[x^{2}]
\end{equation}
where $x$ denotes a random signal, i.e., the EEG signal in one frame.

DE is a generalized form of Shannon's information entropy on continuous variables and can be used to measure the amount of information. DE is defined as:
\begin{equation}
\label{equ_2}
DE=-\int^{+\infty}_{-\infty}p(x)log(p(x))dx
\end{equation}
where $p(x)$ denotes the probability density function of the signal. When the random variables approximately obey the Gaussian distribution $N(\mu,\sigma^{2})$ the DE calculation can be simplified by the following equation: 
\begin{equation}
\label{equ_3}
\begin{aligned}
    DE &= -\int^{+\infty}_{-\infty}\frac{1}{\sqrt{2\sigma^{2}}}e^{-\frac{(x-\mu)^2}{2\sigma^2}}log(\frac{1}{\sqrt{2\sigma^{2}}}e^{-\frac{(x-\mu)^2}{2\sigma^2}})dx \\
       &= \frac{1}{2}log(2\pi e\sigma^2)
\end{aligned}
\end{equation}
where $\mu$ and $\sigma$ denote the mean and standard deviation of the signal $x$ respectively, and $e$ denotes the Euler constant.

Therefore, each sample $x\in\mathbb{R}^{2T\times 2f\times C}$ has three dimensions after feature extraction. Where $T$ represents the time length of the sample, $f$ represents the number of frequency bands, and $C$ represents the number of channels of the EEG data. Finally, z-score normalization is employed for each sample.


\subsection{Spectral attention block}
We extracted the PSD and DE features of the EEG signals at different frequency bands. The EEG signals on different frequency bands reflect different physiological states of human beings. For example, low-frequency EEG signals are often seen when humans are sleeping or resting, while high-frequency EEG signals are usually seen when people are anxious or subject to strong emotional fluctuations\cite{wang2021review}. Therefore, EEG signals can be discriminative for emotion in the frequency domain. To extract the spectral characteristics, we perform cross-band and cross-feature attention calculations on the features extracted on different frequency bands in the spectral attention block. The spectral attention block can be expressed as: 
\begin{equation}
\label{equ_4}
z^{fa}_{t,0}=x_t + E^{fa}_{pos}\quad t=1,2,\cdots,2T
\end{equation}
\begin{equation}
\label{equ_5}
h^{fa}_{t,l}=LN(MHA(z^{fa}_{t,{l-1}})+z^{fa}_{t,{l-1}})
\begin{aligned}
\quad l&=1,2,\cdots,L \\
\quad t&=1,2,\cdots,2T
\end{aligned}
\end{equation}
\begin{equation}
\label{equ_6}
z^{fa}_{t,l}=LN(MLP(h^{fa}_{t,l})+h^{fa}_{t,l})
\begin{aligned}
\quad l&=1,2,\cdots,L \\
\quad t&=1,2,\cdots,2T
\end{aligned}
\end{equation}
where $E^{fa}_{pos}\in\mathbb{R}^{2f\times C}$ denotes the position encoding of the frequency domain features, $t$ denotes the frame in the sample, and $l$ denotes the number of layers. As shown in Fig.1, we employ multi-head attention (MHA)\cite{vaswani2017attention} to the attention calculation of the EEG signal, and then add a multi-layer perceptron (MLP) after this in the transformer encoder. Residual connection\cite{he2016deep} and layer normalization (LN)\cite{ba2016layer} are employed to both MHA and MLP block for accelerating network training. To learn the common features in EEG signals over different time periods, we use the same spectral attention block to train different frames. Meanwhile, this can also greatly reduce the number of parameters of the model. In other words, for different frames in the same sample, the transformer encoders in the spectral attention block share the same parameters.

\subsection{Spatial attention block}
The channel of the EEG signals represents the location of the brain sampled by the electrodes. Similar to EEG frequency characteristics reflecting the different physiological states of humans, each brain region is also responsible for different functions. For example, the frontal part of the cerebral cortex is generally responsible for human physiological emotions\cite{wang2021review}. Correlation between EEG signals from different channels reflects the functional connectivity between different brain regions. Therefore, we perform self-attention calculations on channels to explore the functional connectivity of the brain which contains available information in space.  The spatial attention block can be expressed as:
\begin{equation}
\label{equ_7}
z^{sa}_{t,0}=tran(z^{fa}_{t,L}) + E^{sa}_{pos}\quad t=1,2,\cdots,2T
\end{equation}
\begin{equation}
\label{equ_8}
h^{sa}_{t,l}=LN(MHA(z^{sa}_{t,{l-1}})+z^{sa}_{t,{l-1}})
\begin{aligned}
\quad l&=1,2,\cdots,L \\
\quad t&=1,2,\cdots,2T
\end{aligned}
\end{equation}
\begin{equation}
\label{equ_9}
z^{sa}_{t,l}=LN(MLP(h^{sa}_{t,l})+h^{sa}_{t,l})
\begin{aligned}
\quad l&=1,2,\cdots,L \\
\quad t&=1,2,\cdots,2T
\end{aligned}
\end{equation}
where $E^{sa}_{pos}\in\mathbb{R}^{C\times 2f}$ denotes the position encoding of spatial information and $tran()$ denotes the transpose operation. Similar to the spectral attention block, the transformer encoder is employed and the parameters of the encoders in this block are shared for different frames in the same sample.


\subsection{Temporal attention block}
EEG signals are time-series and acquired by sampling at different times. When humans are stimulated or have emotional fluctuations, it can be reflected in the changes in EEG signals over time. Therefore, EEG signals also carry a large amount of useful information in the time domain. Since emotional fluctuations may only occur at a specific period, not each frame in the sample is critical to the analysis. Hence, in the temporal block, the model assigns an attention score to frames within the sample to reflect the importance of each frame, as shown in Fig. 1. The critical frames are then emphasized and retained by having a large weight in the weighted summation. To calculate the attention score of each frame, the output of the previous spatial attention block is flattened from $Z^{sa}_{L}\in\mathbb{R}^{2T\times 2f\times C}$ to $Z^{ta}\in\mathbb{R}^{2T\times 2fC}$. The computation is as follows: 
\begin{equation}
\label{equ_10}
A_{tem}=Softmax(Z^{ta}W^{T}_{tem}+b_{tem})
\end{equation}
\begin{equation}
\label{equ_11}
O^{ta}=A^{T}_{tem}Z{ta}
\end{equation}
where $W^{T}_{tem}$ and $b_{tem}$ are learnable parameters, and $O^{ta}$ denotes the output of the temporal attention block.

\subsection{Classification layer}
After the raw EEG signals have been passed through the spectral attention block, the spatial attention block, and the temporal attention block, the output is a representation that integrates all the available information on multiple dimensions. In order to fuse the global information of the representation and output a final classification result, a classifier layer is employed. The classification layer is a single layer of a fully connected neural network. After flattening the output of the temporal attention block into a 1D vector, the classification layer is used to obtain the final results, and the whole neural network is optimized with the cross-entropy loss function:
\begin{equation}
\label{equ_12}
L=-\frac{1}{N}\sum^N_{n=1}\sum^C_{c=1}y^c_{n}log(\hat{y}^c_n)
\end{equation}
where $N$ denotes the number of batch sizes and $C$ denotes the number of categories. $y^c_n$ and $\hat{y}^c_n$ are the one-hot label and predicted probability of the corresponding categories, respectively.

\section{Experiment}
\subsection{Dataset}
DEAP is a public dataset for EEG emotion recognition\cite{koelstra2011deap}. 32 Subjects were asked to watch 40 1-minute music videos and record their emotion level of valence and arousal from 1 to 9 based on an online self-assessment. Depending on the above level, we divided the dataset into two classes with a threshold value of 5. The EEG signals were acquired according to the 10/20 system at 512 Hz with 32 channels of EEG. Then the data was downsampled to 128Hz and passed to a filter between 4 and 45 Hz. It is worth noting that each trial consists of 3-second pre-trial baseline and 60-second emotion related signals. Following the previous work\cite{xiao20224d}, we calculated DE features by subtracting baseline DE features from pre-trial signals.

The SEED\cite{zheng2015investigating} and SEED-IV\cite{zheng2018emotionmeter} datasets were collected by the BCMI lab at Shanghai Jiao Tong University and have been widely used in emotion recognition research. The SEED dataset contains three emotions: positive, negative, and neutral. Subjects were asked to watch videos of the three emotions to capture the corresponding EEG signals. A total of 15 subjects participated in the experiment. Each subject watched 15 videos, 5 videos for each emotion, and each video was about 4 minutes long. There was a 45-second self-assessment period and a 15-second break set between videos. The data was collected with the 62-channel ESI NeuroScan System, downsampled to 200Hz and filtered with a bandpass frequency filter from 0-75Hz. 

SEED-IV contains the emotions of happy, sad, neutral, and fear. In the same way, 15 subjects participated in the experiment and were asked to watch corresponding emotional film clips. Each subject's experimental task contained 24 trails, each consisting of a 5-second hint of start, a 2-minute video, and a 45-second self-assessment. As with SEED, the EEG signals were acquired using the ESI NeuroScan System, which consists of 62 channels and downsampled to 200 Hz. After downsampling, a 1-75 Hz bandpass filter was employed to remove noises.
\subsection{Experiment Design}
In order to evaluate our model in the emotion recognition task, we designed the following experiments for a comprehensive comparison. Firstly, we compared AMDET with other current state-of-the-art models. Then, we designed ablation experiments to explore the effect of each part of our model. At last, a visualization experiment was conducted to investigate the characteristic of EEG data. Below is a description of our experiments: 
\subsubsection{Baseline models}
\begin{itemize}
    \item {SVM\cite{suykens1999least}: A generalized linear classifier that solves the maximum margin hyperplane for samples}
    \item {BiHDM\cite{li2018novel}: A recurrent neural network-based model of left- and right-hemisphere differences for EEG emotion recognition}
    \item {RGNN\cite{zhong2020eeg}: A regularized graph neural network considering the biological topology among different brain regions to capture both local and global relations among different EEG channels.}
    \item {4D-CRNN\cite{shen2020eeg}: A convolutional recurrent neural network that extracts spatial, spectral and temporal domain features of EEG signals for emotion recognition.}
    \item {SST-EmotionNet\cite{jia2020sst}: An attention-based two-stream CNN that simultaneously integrates spatial-spectral-temporal features in a single network framework.}
    \item {4D-aNN\cite{xiao20224d}: A 4D attention-based neural network consisting of a CNN and a bidirectional LSTM.}
\end{itemize}
\subsubsection{Ablation Experiment}
In our approach, we customize the model with different structures in each of the three dimensions to capture the abundant features of the EEG signal. In order to investigate the role of each part of the model, we conducted ablation experiments to explore the performance of the model by removing the spectral attention block, the spatial attention block, and the temporal attention block, respectively.

\begin{table*}[!t]
\caption{The performance comparison of the state-of-the-art models. Results are classification accuracy rates (average ± standard deviation)(\%)\label{tab:table1}}
\centering
\setlength{\tabcolsep}{7mm}{
\begin{tabular}{c|c|c|c|c}
\hline
Model          & DEAP-Arousal        & DEAP-Valence        & SEED                & SEED-IV             \\ \hline
SVM            & 89.99±6.74          & 89.33±7.41          & 83.99±9.72          & 56.61±20.05         \\
BiHDM          & -                   & -                   & 93.12±6.06          & 74.35±14.09         \\
RGNN           & -                   & -                   & 94.24±5.95          & 79.37±10.54         \\
4D-CRNN        & 94.58±3.69          & 94.22±2.61          & 94.74±2.32          & -                   \\
SST-EmotionNet & -                   & -                   & 96.02±2.17          & 84.92±6.66          \\
4D-aNN         & 97.39±1.75          & \textbf{96.90±1.65} & 96.25±1.86          & 86.77±7.29          \\
ours           & \textbf{97.48±0.99} & 96.85±1.66          & \textbf{97.17±0.93} & \textbf{87.32±1.79} \\ \hline
\end{tabular}}
\end{table*}
\begin{figure*}[!t]
\centering
\includegraphics[width=7in]{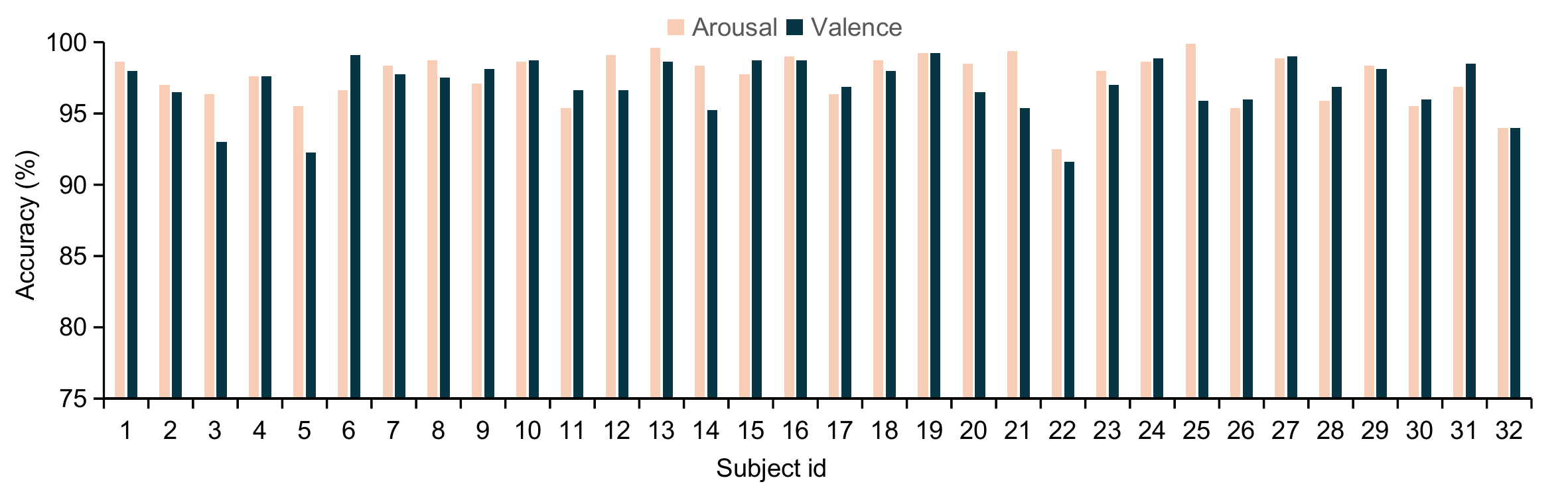}
\caption{The performance of AMDET on DEAP dataset. We conducted the five-fold cross-validation experiment for each subject.}\label{fig:3}
\end{figure*}
\subsubsection{Visualization and EEG channel selection}
We design our model based on attention mechanism which makes the model learn and focus on the significant part in spectral, spatial and temporal domains. As a result, the AMDET achieve state-of-the-art performance. But in the meantime, it is of value to know what the trained model has learned for EEG emotion recognition, i.e., the interpretability of the deep model, for instance, to explore the correlation between emotions and each channel or each frequency band, or to find some specific time domain characteristics of EEG. We adopt Grad-CAM\cite{selvaraju2017grad} to visualize where the model attention is. Grad-CAM (Gradient-weighted Class Activation Mapping) uses gradient to measure the influence of the elements in feature extracted by the model on the prediction results. It is able to highlight the important regions in the image for predicting the concept. 

Different channels in EEG data represent different regions of the cerebral cortex, and different brain regions are responsible for different physiological functions. On the SEED dataset, the number of channels is 62, which were obtained from different brain positions. However, an excessive number of channels not only increases the computational effort but also makes the practical application of brain-computer interaction difficult. Therefore, it is of great importance to reduce the number of channels used when analyzing EEG data. We aim to identify crucial brain regions or channels for emotion recognition. Based on this, we further made a selection of EEG channels. We reduced the number of EEG channels used for model training, from 62 to 32, 16, and 8 channels, respectively, and discussed the effects of the number of channels on the recognition performance.
\subsection{Experiment Detail}
All experiments in this paper were conducted on an NVIDIA TITAN Xp GPU. The number of layers in the spectral attention block and the number of layers in the spatial attention block are set to 1 and 1, respectively. The number of heads in the spectral attention block and the number of heads in the spatial attention block are set to 2 and 2, respectively. The number of frequency bands are set to 4 for DEAP dataset (theta[4-8Hz], alpha[8-14Hz], beta[14-31Hz], gamma1[31-50Hz]) and set to 5 for SEED dataset and SEED-IV dataset (theta[4-8Hz], alpha[8-14Hz], beta[14-31Hz], gamma1[31-50Hz], gamma2[50-75Hz]). We use the AdamW optimizer with learning rate, weight decay, and batch size of 1e-3, 1e-6, and 16, respectively, to optimize the neural network. For DEAP dataset, we used only DE feature. We conducted experiments on each subject. For SEED and SEED-IV datasets, we calculated the average accuracy of each subject in the 3 experiments. We used five-fold cross-validation for all experiments.
\section{Results and Discussion}
\begin{figure*}[!t]
\centering
\includegraphics[width=7in]{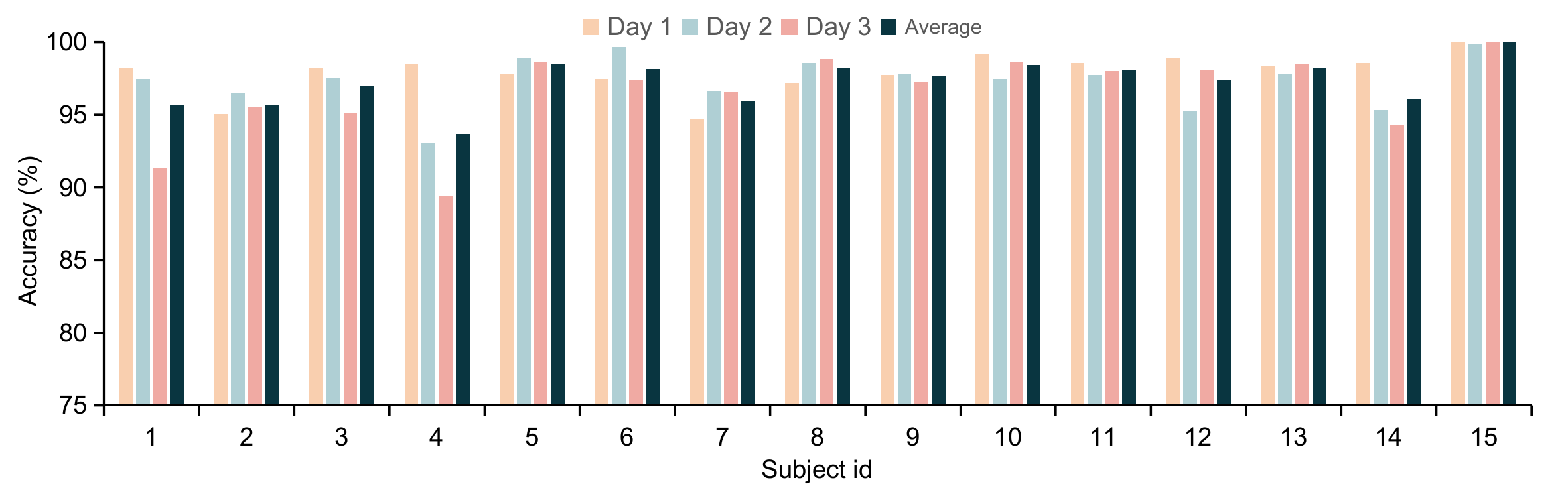}
\caption{The performance of AMDET on SEED dataset. We conducted the five-fold cross-validation experiment for each subject and calculated the average results for different days.}\label{fig:4}
\end{figure*}
\begin{figure*}[!t]
\centering
\includegraphics[width=7in]{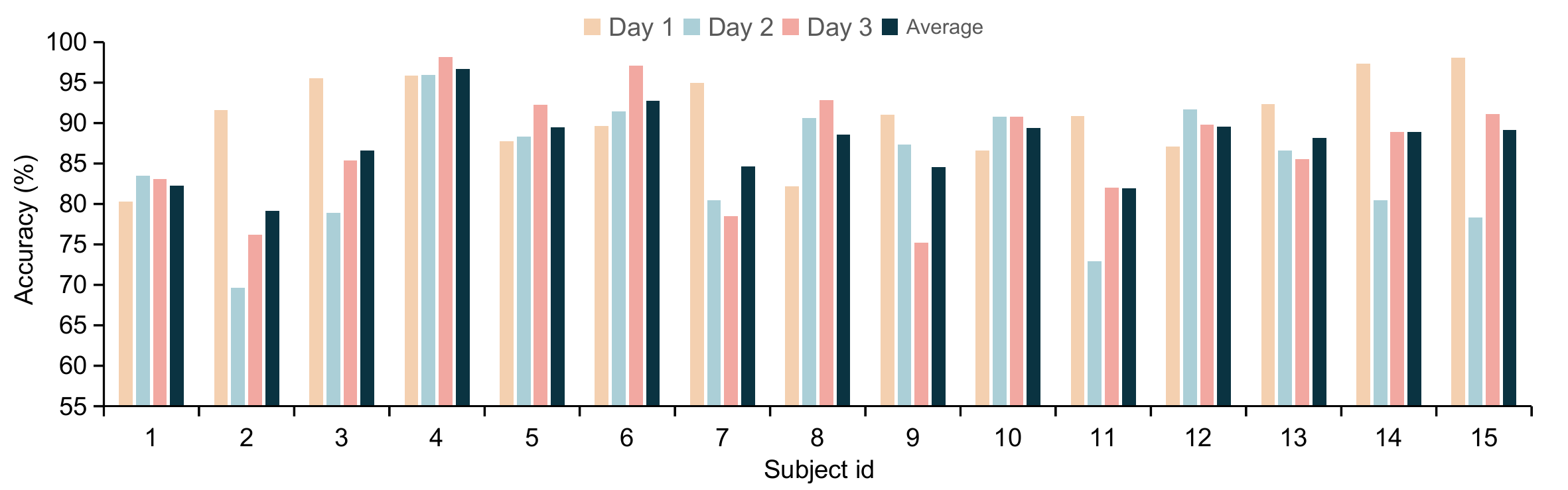}
\caption{The performance of AMDET on SEED-IV dataset. We conducted the five-fold cross-validation experiment for each subject and calculated the average results for different days.}\label{fig:5}
\end{figure*}
\subsection{Compared with baseline}
In order to compare our model with baseline models, we conducted experiments on DEAP-Arousal, DEAP-Valence, SEED, and SEED-IV datasets. Table \Rmnum{1} shows the experimental results. The experimental results show that the deep learning methods are generally better than the traditional machine learning methods, and the accuracy of SVM is only 89.33\%, 89.99\%, 83.99\%, and 56.61\%. RGNN and BiHDM explore the spatial properties of EEG signals and achieve 94.24\%/79.37\% and 93.12\%/74.35\% accuracy on SEED and SEED-IV datasets respectively. 4D-CRNN does not just focus on the features in the spatial domain, but extracts the spectral-spatial-temporal features in EEG by CNN and RNN, reaching a better accuracy of 94.22\% on SEED datasets. In addition, it achieves 94.22\% and 94.58\% accuracy on DEAP datasets. SST-EmotionNet and 4D-aNN tried to integrate the attention mechanism into their models in combination with CNN and LSTM. They also fused the features of EEG signals on all domains, and finally achieved 96.02\%/84.92\% and 96.25\%/86.77\% results on SEED and SEED-IV datasets, respectively. Our proposed model utilizes a transformer-based method to extract the frequency and spatial features of the EEG signal, then uses a temporal attention block to help the model focus on significant frames. The final result outperforms all baseline models and reaches 96.85\%, 97.48\%, 97.17\% and 87.32\% on four datasets. It is worth noting that the results of the approaches focusing on multiple domains are superior to those that study only a single domain, which illustrates the value of exploring the multi-dimensions characteristics of the EEG signal. At the same time, the comparison with the similar attention-based models SST-EmotionNet and 4D-aNN indicates that Transformer is more appropriate than CNN and LSTM for detecting critical and discriminative features on different domains.

AMDET also has the lowest standard deviation compared to baseline models, which means it is more adaptable to different people. Fig. 3, Fig. 4, and Fig. 5 demonstrate each subject experiment results individually on DEAP, SEED, and SEED-IV datasets. For the DEAP dataset, there are total 32 subjects and 2 experiments, arousal and valence classification, almost all of the accuracy achieve above 95\%, except for subject 5, 22, and 32. Their accuracy is 95.5\%/92.25\%, 92.5\%/91.625\%, and 94\%/94\% on arousal and valence classification. The SEED dataset includes 15 subjects, and each subject has three days of experimental data. For classification on SEED dataset, there are 6 subjects, subject 1, 2, 3, 4, 7, and 14, performed below the average accuracy of 97\%. As for SEED-IV dataset, it includes 15 subjects and 3 days data as SEED dataset, and has an extra emotion fear for classification. For the classification on SEED-IV dataset, there are 5 subjects whose accuracy is below 85\%, and they are subject 1, 2, 7, 9, and 11. In the SEED-IV dataset, the accuracy for the three days varied greatly, with the 2nd day's accuracy rate being about 6\% lower than the 1st day's and the 3rd day's being about 4\% lower. There may be some accidents during the data collection on the 2nd and 3rd days. 

Experiment results prove that amdet can achieve excellent performance of higher accuracy and lower standard deviation on DEAP, SEED, and SEED-IV datasets. It also indicates the effectiveness and necessity of fusing multiple dimensional information of EEG for classification.
 
\begin{figure}[ht]
\centering
\includegraphics[width=3.4in]{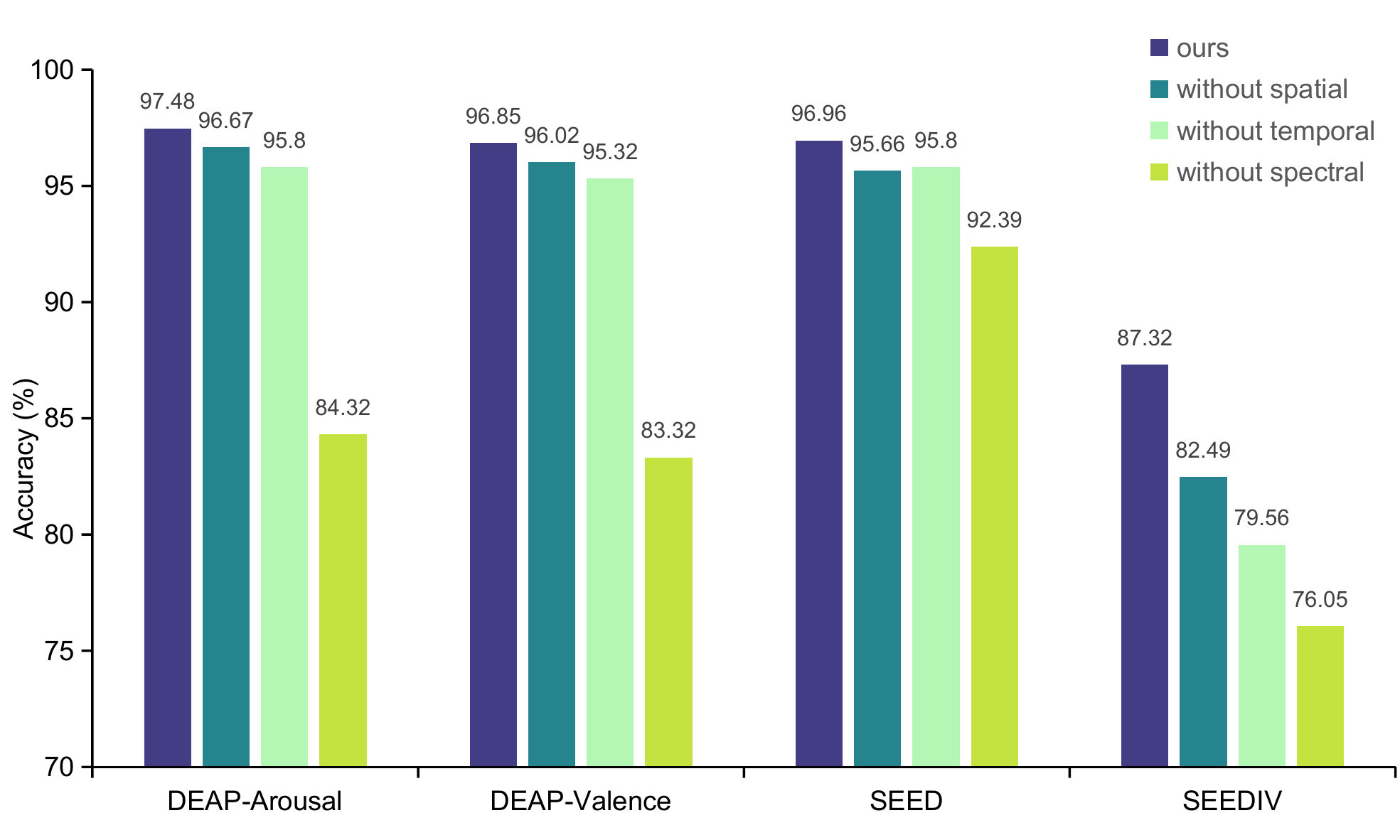}
\caption{Results of the ablation study on DEAP-Arousal, DEAP-Valence, SEED, and SEED-IV datasets. We remove the spectral transformer block, the spatial transformer block, and the temporal attention block separately to investigate the role of each module.}\label{fig:6}
\end{figure}
\subsection{Ablation Study}
In our method, the model has three blocks for feature extraction of EEG signals, which are used to calculate attention in three different dimensions. To explore the role of different blocks of our model in the classification, we remove the spectral attention block, the spatial attention block and the temporal attention block, respectively, and reserve only the remaining two blocks. The results are shown in Fig. 6. We found that the spectral layer is more important than the spatial and temporal layers. For arousal and valence classification on the DEAP dataset, the performance of the model decreased significantly when the spectral attention block is removed, by 13.53\% and 13.16\%, respectively. In comparison, the model accuracy dropped slightly when the spatial attention block is removed, by 0.83\% and 0.81\%, respectively. Similarly, for the SEED and SEED-IV datasets, the accuracy decreased the most when the frequency attention layer was removed, by 4.57\% and 11.28\%, respectively. The spatial attention layer affected the accuracy the least on the SEED-IV dataset, with only a 4.83\% decrease after removing. On the SEED dataset, the impact of the spatial attention block and the temporal attention block were not significantly different after removal, decreasing by 1.3\% and 1.16\%, respectively. Therefore, we consider the features on different frequency bands to be the most important for the emotion recognition task. In other words, the model mainly focuses on the frequency domain features to classify different emotions. After extracting DE and PSD features, EEG signals are more discriminative. The spatial and temporal features of EEG data, on the other hand, have less importance in emotion recognition compared to the frequency domain features and do not play a decisive role in emotion recognition.
\begin{figure}[ht]
\centering
\includegraphics[width=3.4in]{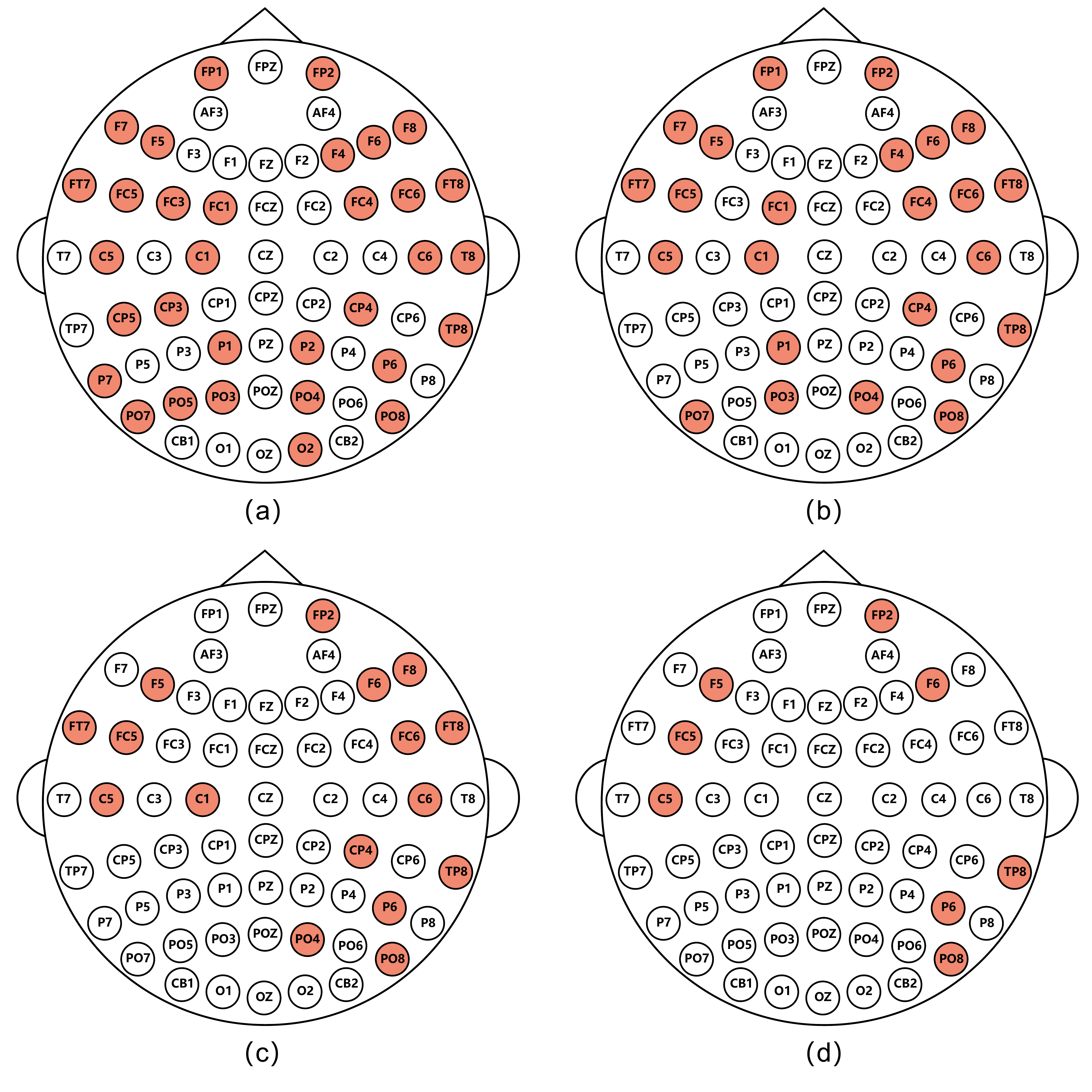}
\caption{Top 32, 24, 16, 8 channels (62 in total) according to visualization for subject 8th on SEED dataset. We rank the channels based on the heat map and select the ones ranked at the top.}\label{fig:7}
\end{figure}
\subsection{EEG channel visualization and channel selection}
To explore what the trained model learns, we adopt Grad-CAM\cite{selvaraju2017grad} to visualize the concern of the model. We use the features map and gradient to generate a heatmap that shows the import part having a greater impact on the prediction.

In order to identify the critical channels for emotion recognition, we need to investigate the influence of each channel on the prediction of the trained model. After training the model with all 62 channels, we employed Grad-CAM for visualization. The channels were ranked according to their weights for emotion classification and those with the greater weights were selected for subsequent experiments. Fig. 7 shows the top 32, top 24, top 16 and top 8 channels for the 8th subject on the SEED datasets. It can be seen that the top 8 channels are P6, C5, TP8, F6, FP2, PO8, FC5 and F5, which are roughly concentrated in the temporal lobe and the parietal lobe in terms of cerebral location. The result shows that these regions of the cerebral cortex have a greater effect on the results of emotion recognition than the other parts.
\begin{figure}[ht]
\centering
\includegraphics[width=3.4in]{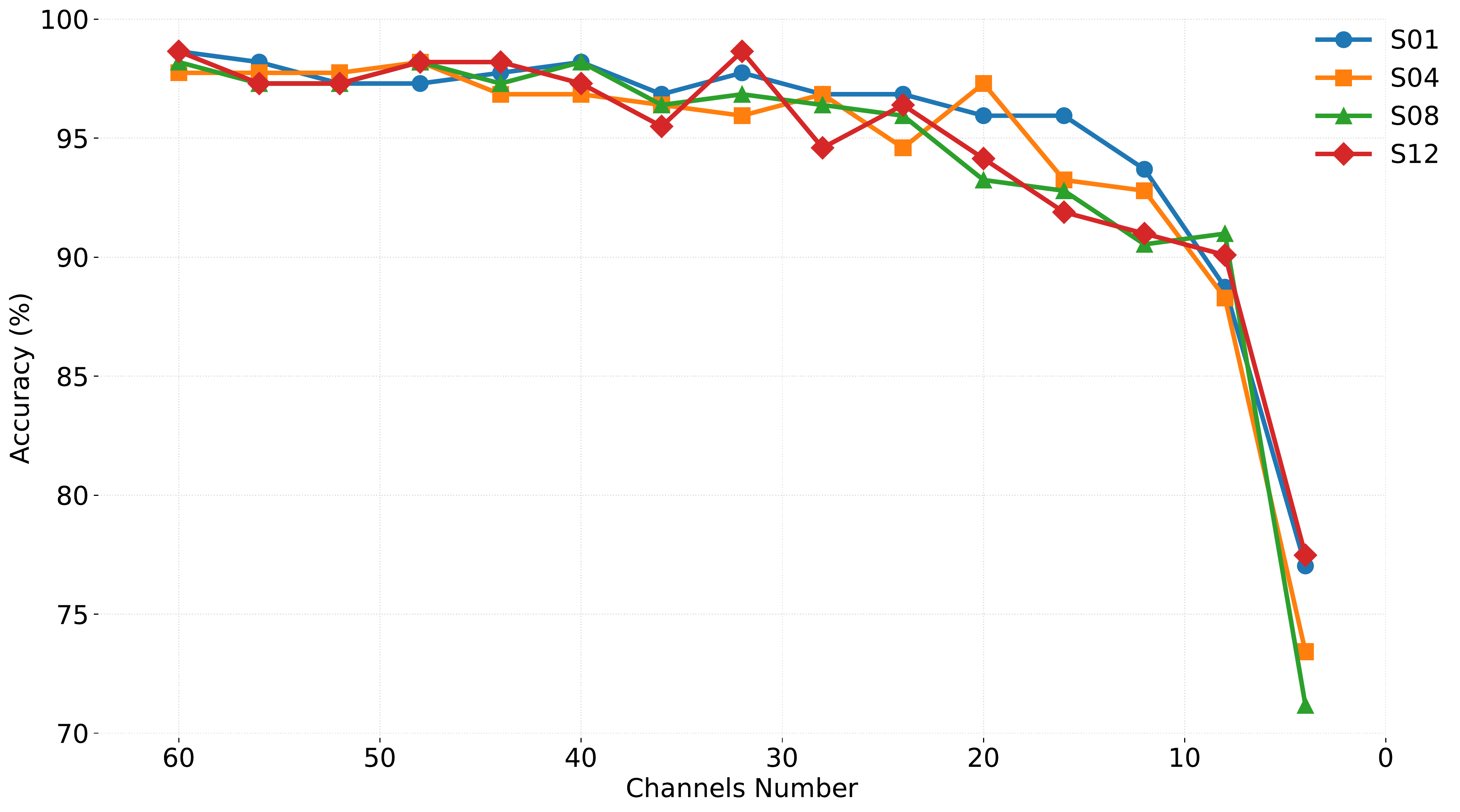}
\caption{Results of channel reduction experiment of subject 1st, subject 4th, subject 8th and subject 12th on SEED dataset. We reduce the number of channels of the input tensor from 62 channels to 2 channels in order of the heat map.}\label{fig:8}
\end{figure}

In addition to the channel visualization of EEG signals, we further investigated the impact of reducing the number of channels. We believe that there may be too much redundant information between EEG channels. On the one hand, reducing the number of channels can make the calculation time shorter; on the other hand, redundant channels for the emotion recognition task can produce noise effects. Therefore, we conducted EEG channel reduction experiments based on previous channel visualization results. We reduced the number of channels sequentially starting from 62 with a stride of 4. The experiment results are shown in Fig. 8. After selecting the 32 channels with the highest importance weights, the accuracy of emotion recognition decreased by only 1\% compared to that of all channels used. As a result, it is clear that these selected 32 channels contain the majority of the information needed for the emotion recognition task, while the remaining channels reflect human emotion less and have an inconsiderable effect on the final task. When the number of channels was reduced to 24, 16, and 8, respectively, the accuracy of emotion recognition started to decrease gradually, by 2\%, 3\%, and 10\%. It can be seen that even when the number of channels is reduced to 8, our model could still achieve an accuracy of about 90\%. However, reducing the number of channels to less than 8 had a substantial impact on the task, yielding a large decrease in accuracy. It would lower the cost of time and computation when the number of channels is reduced. The number of parameters and FLOPs are 0.30M and 0.03G when using 62 channels, while that of 8 channels are only 0.078M and 0.0039G. In addition, fewer channels in inference are of great importance and use for putting EEG-based applications into reality. It means a smaller and more portable collection device. 

\section{Conclusion}
In this paper, we propose a transformer-based model, namely AMDET, for EEG emotion recognition. AMDET achieved state-of-the-art results by extracting and fusing temporal-spatial-frequency features in the EEG signal. Without CNN or RNN enhancing the transformer model, our model is based on a self-attention mechanism, which illustrates the potential for transformers in EEG pattern recognition tasks. The results of the ablation experiments show that information in all three domains is necessary to obtain favorable results on the EEG task, while the information in the frequency domain is of particular significance. Finally, we conduct a channel reduction experiment that selects the channels that contribute the most to the results by visualizing the focus of the model. This reduces the computational effort while ensuring the accuracy of the recognition. On the one hand, the experiment results demonstrate the strong feature extraction ability of our model, which has excellent performance even with few channels in the EEG, and on the other hand, this may also indicate the large redundancy of the EEG signal in the channel dimension. Currently, the visualization is implemented based on Grad-CAM, yet deep learning visualization methods that are more appropriate for EEG need to be devoted to more research, which could explore the role of different channels on different EEG paradigms. The improvement in visualization methods will not only make EEG devices more portable by reducing the number of electrodes but also has the potential to contribute to the development of neuroscience.

\bibliographystyle{IEEEtran}
\bibliography{reference}


 





\end{document}